\documentclass[conference]{IEEEtran}

\usepackage{textcomp}
\usepackage{stfloats}
\usepackage{url}
\usepackage{pifont}
\newcommand{\cmark}{\textcolor{green!50!black}{\ding{51}}} 
\newcommand{\xmark}{\textcolor{red!70!black}{\ding{55}}}   
\usepackage{flushend}
\usepackage{verbatim}
\usepackage{graphicx}
\usepackage{diagbox}
\usepackage{balance}
\PassOptionsToPackage{hidelinks}{hyperref}
\usepackage{hyperref}
\usepackage{makecell}

\usepackage{tabularx}
\newcolumntype{L}{>{\raggedright\arraybackslash}X}%
\newcolumntype{R}{>{\raggedleft\arraybackslash}X}%
\newcolumntype{C}{>{\centering\arraybackslash}X}%
\usepackage{paralist}
\usepackage{lscape}
\usepackage[table]{xcolor}
\usepackage{threeparttable}
\usepackage{paralist}

\usepackage{textcomp}
\usepackage{booktabs}
\usepackage{multirow}
\usepackage{graphicx}
\usepackage[normalem]{ulem}
\useunder{\uline}{\ul}{}

\usepackage{paralist}
\usepackage{lscape}
\usepackage[table]{xcolor}
\usepackage{threeparttable} 
\usepackage{listings}
\hypersetup{
pdftitle={Improving LLM-Driven Test Generation by Learning from Mocking Information},
pdfauthor={Jamie Lee; Flynn Teh; Hengcheng Zhu; Mengzhen Li; Mattia Fazzini; Valerio Terragni},
pdfsubject={This paper introduces MOCKMILL, an LLM-based approach that improves automated unit test generation by leveraging developer-written mocking information extracted from existing tests.},
pdfkeywords={test generation; LLMs; test doubles; mocking}
}

\usepackage{orcidlink}

\usepackage{rotating}

\usepackage{array} 
\usepackage{booktabs}
\usepackage{microtype}
\usepackage{algpseudocode}
\usepackage{xspace}
\newcommand{\gptfouromini}{\textsc{GPT--4o Mini}\@\xspace}
\newcommand{\gptfivemini}{\textsc{GPT--5 Mini}\@\xspace}
\newcommand{\gptfive}{\textsc{GPT--5}\@\xspace}
\newcommand{\claude}{\textsc{Claude Sonnet 4.5}\@\xspace}
\newcommand{\mockito}{\textsc{Mockito}\@\xspace}
\newcommand{\java}{\textsc{Java}\@\xspace}
\newcommand{\randoop}{\textsc{Randoop}\@\xspace}
\newcommand{\evosuite}{\textsc{Evosuite}\@\xspace}
\newcommand{\maven}{\textsc{Maven}\@\xspace}
\newcommand{\gradle}{\textsc{Gradle}\@\xspace}
\newcommand{\junit}{\textsc{JUnit}\@\xspace}

\newcommand{\tool}{\textsc{MockMill}\@\xspace}

\usepackage[dvipsnames]{xcolor}

\usepackage{eso-pic}
\usepackage{hyperref}

\AddToShipoutPictureFG*{%
\AtPageLowerLeft{%
\raisebox{1.6cm}{%
\hspace*{3.0cm}%
\parbox{\dimexpr\paperwidth-5.0cm\relax}{
This is the authors’ version of the paper published in IEEE International Conference on
Software Testing and Verification Workshops (ICSTW) - International Workshop on Artificial Intelligence in Software Testing (AIST).}%
}%
}%
}

\usepackage{refcount}
\usepackage{tcolorbox}
\newcounter{rq}

\begin{document}

\title{Improving LLM-Driven Test Generation  \\ by Learning from Mocking Information}

\author{
\IEEEauthorblockN{Jamie Lee\IEEEauthorrefmark{1}, Flynn Teh\IEEEauthorrefmark{1}, Hengcheng Zhu\orcidlink{0000-0002-3082-5957}\IEEEauthorrefmark{2}, Mengzhen Li\orcidlink{0000-0002-6728-1375}\IEEEauthorrefmark{3}, Mattia Fazzini\orcidlink{0000-0002-1412-1546}\IEEEauthorrefmark{3}, Valerio Terragni\orcidlink{0000-0001-5885-9297}\IEEEauthorrefmark{1}}
\IEEEauthorblockA{
\IEEEauthorrefmark{1}\textit{University of Auckland}, Auckland, New Zealand\\
\IEEEauthorrefmark{2}\textit{The Hong Kong University of Science and Technology}, Hong Kong SAR\\
\IEEEauthorrefmark{3}\textit{University of Minnesota}, Minneapolis, USA
}
Emails: \IEEEauthorrefmark{1}\{eejl773, fteh492, vter674\}@aucklanduni.ac.nz, \IEEEauthorrefmark{2}hzhuaq@connect.ust.hk, \IEEEauthorrefmark{3}\{li001618, mfazzini\}@umn.edu
}

\maketitle

\begin{abstract}
    Large Language Models (LLMs) have recently shown strong potential for automated unit test generation. This has motivated us to investigate whether developer-defined test doubles (commonly referred to as mocks) available in existing test suites can be leveraged to improve LLM-driven test generation. To this end, we propose \tool, an LLM-based technique and tool that generates test cases by exploiting mocking information automatically extracted from developer-written tests. \tool targets components that are replaced by test doubles in existing tests and uses the encoded stubbings and interaction expectations to guide test generation, combined with an iterative generation-and-repair process to ensure executable tests.
    We evaluated \tool on 10 open-source classes from six \java projects using four LLMs, and compared the generated tests with existing project tests and tests produced by baseline approaches. The results show that \tool's tests cover lines of code and kill mutants that existing tests and baseline-generated tests miss. Overall, our findings provide preliminary evidence that leveraging mocking information is a complementary and effective way to enhance LLM-based test generation.
\end{abstract}


\section{Introduction}

Software underpins modern life, making rigorous testing essential to ensure correctness~\cite{myers2011art}. 
Yet writing high-quality tests is labor-intensive~\cite{bertolino2007software}, motivating decades of automated test generation research~\cite{bertolino2007software,anand2013orchestrated}. Long-standing approaches range from random testing to search-based methods~\cite{jahangirova2023sbfttrack}. Recently, the emergence of Large Language Models (LLMs) has opened a new frontier for automated test generation~\cite{
bodicoat2025understanding,ouedraogo2024large,terragni2025future}. LLMs, trained on vast code corpora, can produce useful tests, often with minimal human guidance~\cite{schafer2023empirical,wang2024software}. This capability has sparked interest in LLM-based testing~\cite{wang2024software}. However, while the results of LLM-driven test generation are highly promising, many aspects of LLM-based test generation remain underexplored, particularly the impact of providing specific auxiliary artifacts beyond the class under test~\cite{wang2024software,schafer2023empirical}.

In this paper, we explore how to improve LLM-driven test generation by leveraging \emph{test doubles}~\cite{meszaros2007xunit} (or more informally also referred to as \emph{mocks}). Test doubles are test artifacts commonly used to manage the complexity of interacting components during testing. They are lightweight stand-in objects that mimic real components, allowing developers to isolate the software under test from its dependencies~\cite{meszaros2007xunit}, also referred to as \emph{mocked components}. Test doubles can be configured to return predefined responses for specific inputs (\emph{stubbings}) or to check expected interactions (\emph{verify operations}). Through them, developers implicitly specify how dependent components should behave.
Test doubles are central to modern software testing~\cite{meszaros2007xunit}, and studies across programming languages show that stubbing and verify operations are widely used in both industry and open-source software~\cite{mostafa2014empirical,spadini2019mock,2022_icse_fazzini_use,zhu2025understanding}.
Tests generated from test double information can help verify dependent components when they are part of the software system.

Existing LLM-based approaches typically generate tests using the code under test and its associated documentation or comments~\cite{schafer2023empirical,wang2024software}, making the rich usage patterns and expected behaviors encoded in test doubles an untapped resource for improving automated test generation. To bridge this gap, we introduce \textbf{\tool}, an \emph{LLM-based approach that leverages the mocking information from an existing test suite to generate new tests}. Given the software under test and its test suite as inputs, \tool generates tests for those components that are replaced by test doubles in existing tests. The approach starts by automatically extracting mocking information from the test suite via static analysis. \tool then employs an LLM-guided approach to generate tests for targeted components. The approach integrates a generation-and-repair loop: the extracted mock information is incorporated into the LLM input to generate candidate tests, and any compilation or runtime issues in the tests are iteratively fixed via LLM-based corrections. 
By guiding an LLM with this existing, developer-defined information, \tool produces tests that align with the behavioral expectations already encoded in the test suite.

We implemented \tool and evaluated it on 10 classes from six open-source \java projects using four LLMs (\gptfouromini, \gptfivemini, \gptfive, and \claude). We compare the tests generated by \tool with existing project tests and with two baselines: an LLM-based approach that does not utilize mocking information and random test generation (\randoop)~\cite{pacheco2007randoop}. We measured line and mutation coverage to assess effectiveness. The results suggest that \tool covers code and kills mutants missed by existing and baseline tests, providing preliminary evidence that it complements existing or automatically generated tests. We released \tool's code and experimental data~\cite{mockmill-replication}.

\section{Intuition and Running Example}

Our key \textbf{intuition} is that \emph{mocking information encodes meaningful knowledge about how software components should be used and this can be leveraged to create useful tests}. This work specifically leverages \emph{stubbings} and \emph{verify operations}~\cite{mockito_doc} as \emph{mocking information} (or \emph{mocking data}), which often specify: \begin{inparaenum}[(i)]
	\item which methods of dependent components are invoked,
	\item the arguments passed to the methods, and
	\item the return values or exceptions resulting from those invocations.
\end{inparaenum}
Stubbings are the natural way to encode (i), (ii), and (iii). Furthermore, verify operations can also specify (i) and (ii) by checking that certain methods have been called with certain arguments during test execution~\cite{mockito_doc}. Such structured behavioral information serves as \emph{implicit usage documentation of dependent components}. We use \emph{dependent components}, \emph{replaced components}, \emph{mocked components} interchangeably to refer to those components that are replaced by test doubles and on which stubbings and verify operations are defined. When provided to an LLM during test generation, the information can guide the model toward realistic and meaningful tests that improve the coverage of the test suite. \tool exploits this insight by extracting mocking information from existing tests and using the information to generate tests for the mocked components. These components are the \emph{target components} in the test generation process and are the dependent components in existing tests.

\noindent
Figure~\ref{lst:running-example} shows a \textbf{running example}, illustrating how \tool leverages mocking information and the advantage of incorporating it. 
The figure includes three tests: the original developer-written test {\small{\texttt{pageQueryAopLogsTest}}} (from which \tool extracts the mocking data), the test generated by an LLM-driven baseline without providing mocking information ({\small{\texttt{baseline\_test}}}), and the test generated by \tool using the extracted mocking data ({\small{\texttt{mockmill\_test}}}). {\small{\texttt{pageQueryAopLogsTest}}} contains mocking data for the {\small{\texttt{AopLogRepository}}} class (i.e., the dependent component). This component is the target component (i.e., target class) for {\small{\texttt{baseline\_test}}} and {\small{\texttt{mockmill\_test}}}. {\small\texttt{AopLogRepository}} is also a subject of our evaluation (Section~\ref{sec:eval}).


The baseline approach generated a test that exercises only the {\small{\texttt{fetchBy}}} method, representing the most common and straightforward usage of the {\small{\texttt{repository}}} object. The existing developer test {\small{\texttt{pageQueryAopLogsTest}}} mocks {\small{\texttt{AopLogRepository}}} and stubs its {\small{\texttt{pageFetchBy}}} method, which reflects a more advanced usage scenario involving paging. 
Providing the mocking information in {\small{\texttt{pageQueryAopLogsTest}}} enables \tool to guide an LLM to generate a test that exercises the paging mechanism of {\small{\texttt{AopLogRepository}}}, thereby covering more lines of code and killing mutants injected in that method.

\begin{figure}
    \centering
\begin{lstlisting}[language=java,basicstyle=\ttfamily\scriptsize]
// Developer-written test with mocking info
@Test
void pageQueryAopLogsTest() {
  ... aopLogRepository = mock(AopLogRepository.class);
  PageRequestDto pageDto = PageRequestDto.of(1, 10);
  AopLogQueryDto queryDto = new AopLogQueryDto();
  when(aopLogRepository.pageFetchBy(pageDto, queryDto)).thenReturn(mockResult);
  // ...}
(*@\noindent\textcolor{black}{\rule{\linewidth}{0.4pt}}@*)
// Baseline-generated test without mocking info (line 7)
@Test
void baseline_test() {
  AopLogRepository repository = new AopLogRepository();
  AopLog aopLog = new AopLog();
  aopLog.setId(1L);
  repository.insert(aopLog);
  AopLogQueryDto queryDto = mock(AopLogQueryDto.class);
  when(queryDto.getId()).thenReturn(2L);
  List<AopLog> result = repository.fetchBy(queryDto);
  assertTrue(result.isEmpty());}
(*@\noindent\textcolor{black}{\rule{\linewidth}{0.4pt}}@*)
// MockMill-generated test giving mocking info (line 7)
@Test
void mockmill_test() {
  AopLogRepository aopLogRepository = new AopLogRepository();
  PageRequestDto pageRequestDto = PageRequestDto.of(0,10);
  AopLogQueryDto queryDto = new AopLogQueryDto();
  Result<Record> mockResult = mock(Result.class);
  Result<Record> result = aopLogRepository.pageFetchBy(pageRequestDto, queryDto);
  assertEquals(mockResult, result);}
\end{lstlisting}
\vspace{0pt}
    \caption{Running example based on the \texttt{AopLogRepository} class.}
    \vspace{0pt}
    \label{lst:running-example}
\end{figure}

\section{\tool}

\begin{figure*}
    \centering
    \includegraphics[width=0.88\linewidth]{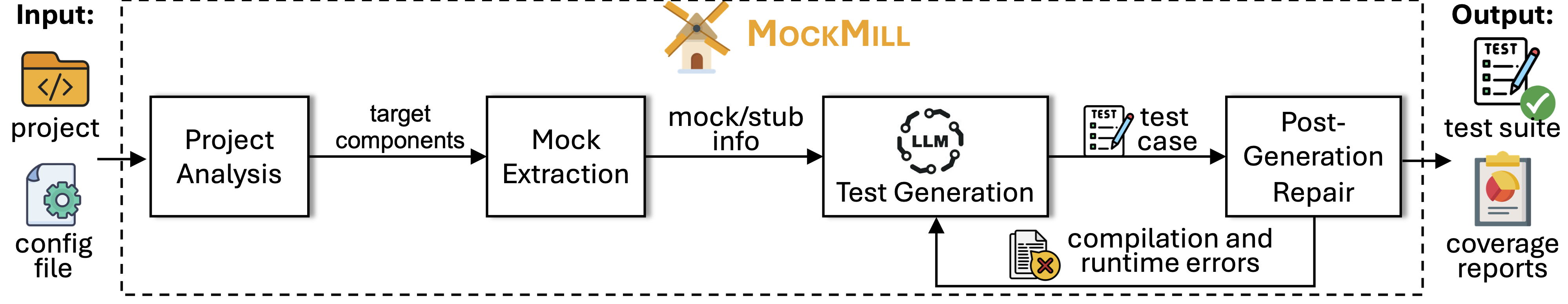}
    \vspace{0pt}
    \caption{High-level overview of \tool's workflow.}
     \vspace{0pt}
    \label{fig:tool}
\end{figure*}

\tool is an LLM-based automated test generation approach that leverages mocking information to generate test cases.
Its methodology comprises four phases: (1) project analysis, (2) \textbf{mock extraction} (introduced in this work), (3) test generation, and (4) post-generation repair.
Phases (1), (3), and (4) follow common patterns in prior LLM-assisted test generation approaches~\cite{schafer2023empirical,wang2024software}, while phase (2) is novel and extracts realistic usage information about mocked components that prior approaches do not explicitly leverage.

Figure~\ref{fig:tool} provides a high-level overview of \tool. \tool takes as input a software project configured with a standard build automation tool (e.g., \maven\ or \gradle\ for \java).  It also accepts a configuration file to filter target components (e.g., classes in \java), specify LLM settings, define the number of repair attempts, and set token limits. \tool begins with the {\small\textsf{Project Analysis}} phase, which finds existing tests using stubbings or verify operations and identifies associated components, which are the target components for test generation.
The {\small\textsf{Mock Extraction}} phase then gathers structural data (i.e., the specific content of stubbings and verify operations) about the dependent component.
This data is passed to the {\small\textsf{Test Generation}} phase, where the LLM produces candidate test cases based on extracted mock data.
Finally, the {\small\textsf{Post-Generation Repair}} phase compiles, executes, and iteratively repairs the generated tests until they run successfully. The output of \tool consists of automatically generated tests for components that are substituted by test doubles in other tests. The output also includes logs about compilation and repair attempts, along with code and mutation coverage reports that help assess the quality of the generated tests. The rest of this section describes \tool's phases in detail. The LLM prompts used by the approach are omitted due to space limitations but are available in the replication package~\cite{mockmill-replication}.

\subsection{Project Analysis}
The {\small\textsf{Project Analysis}} phase identifies the dependent components that can be targets for test generation by analyzing the test code in the project under analysis.
Specifically, this phase parses the abstract syntax tree (AST) of the test files in the project to identify the components that have been replaced by test doubles and that use stubbings or verify operations.
In our implementation, we identified the test cases written using \junit{} 5 and the test doubles created with \mockito{}, since they are the most popular testing and mocking frameworks in the Java ecosystem~\cite{zhu2025understanding}.
Once \tool identifies this information, this phase filters out components that are not part of the project (i.e., components from third-party libraries) by keeping only the classes that are defined in the source code of the project (and not in the libraries used by the project).
The approach performs this step as generated tests should focus on code that belongs to the project.
\tool encodes the relevant targets in a structured representation (i.e., a \texttt{JSON} file) that is then passed to the subsequent phase of the approach.

\noindent
\textbf{Running Example:} 
For the project in Figure~\ref{lst:running-example} (lines 2--9), this phase identifies that the test {\small{\texttt{pageQueryAopLogsTest}}} mocks {\small{\texttt{AopLogRepository}}} and defines a stubbing on it, selecting the class as a candidate for mock-informed test generation.

\subsection{Mock Extraction}
The {\small\textsf{Mock Extraction}} phase focuses on collecting the mocking information only relevant to each target component and structuring it into an intermediate JSON representation (which will be provided to the LLM for test generation). This phase identifies calls to mocking APIs of the supported mocking frameworks (e.g., {\small{\texttt{when(...).then(...)}}} or {\small{\texttt{verify()}}}  in \mockito), along with argument values associated with the calls to the API methods by parsing the AST of the test files. The analysis traverses both test and setup methods to record field assignments, verification calls, and stubbing logic associated with the selected class. 
This focused extraction ensures that the {\small\textsf{Test Generation}} phase receives precise input data without including the entire source files of the test, which could otherwise overload the LLM context window and add noise, reducing the focus and overall quality of generated tests.

\noindent
\textbf{Running Example:} 
In Figure~\ref{lst:running-example}, the {\small\textsf{Mock Extraction}} phase analyzes {\small{\texttt{pageQueryAopLogsTest}}} and records that the method {\small{\texttt{pageFetchBy(pageDto, queryDto)}}} of {\small{\texttt{AopLogRepository}}} is stubbed. 
This extracted detail becomes part of the structured input provided to the generator, ensuring that the LLM is aware of the paging behavior exercised in the developer-written test.

\subsection{Test Generation}

The {\small\textsf{Test Generation}} phase constructs LLM prompts for automated test creation based on the target component and the extracted mock information.
In this phase, a structured prompt is assembled that provides the LLM with (i) explicit test generation instructions, (ii) the complete source code of the target component (e.g., the source code of the target class), and (iii) the data associated with the relevant stubbings and verify operations.
Including the full class source code enables the LLM to reason about the implementation context, dependencies, and behaviors, thereby reducing the likelihood of incorrect assumptions or hallucinations~\cite{bodicoat2025understanding,ouedraogo2024large}.

\noindent
\textbf{Prompt design:} 
The test generation prompt targets mutation-based adequacy~\cite{papadakis2019mutation}, as mutation coverage is widely recognized as one of the most comprehensive test adequacy criteria~\cite{papadakis2019mutation}.
The prompt instructs the LLM to (i) use exact values from stubbings and verify operations; 
(ii) create precise assertions that fail under realistic mutations (bugs); 
(iii) test both true and false execution paths and boundary conditions derived from mocking information; and 
(iv) generate tests that instantiate and exercise real objects of the CUT rather than substituting them with test doubles.

\noindent\textbf{Baselines}. The approach also uses a few-shot prompting strategy~\cite{brown2020language} and provides two examples for the requested task. The examples are based on simple, self-contained \java classes along with corresponding structured mock and stub data extracted from representative tests. This data captures real interaction patterns that reflect how the CUT is used. The examples then include executable unit tests that instantiate the CUT, exercise its behavior, and assert on expected outputs. We intentionally designed the two few-shot examples to be domain-agnostic (and are not based on any of the projects considered in the evaluation). The prompt also does not target or identify specific mutants to kill~\cite{foster2025mutation}. It encourages the LLM to produce tests that satisfy mutation-based adequacy criteria in a general sense.
Finally, the prompt specifies the required output format, directing the LLM to produce a compilable test file that includes all necessary package declarations and import statements.

\subsection{Post-Generation Repair}

After generation, \tool iteratively compiles and repairs tests, following prior LLM-driven test generation~\cite{ravi2025llmloop,schafer2023empirical}. If compilation fails, the resulting error messages are passed to the LLM with targeted repair instructions to produce a corrected test file. This cycle repeats until the tests compile successfully or the retry limit is reached. The same iterative process applies to runtime failures. Once compilation succeeds, \tool executes the tests and, upon failure, provides the error logs to the LLM for correction. This loop continues until all tests pass or the maximum retry limit is reached.

This design assumes a regression testing scenario~\cite{shamshiri2015automatically,terragni-fse-2020}, where failing generated tests are attributed to issues in the tests themselves rather than bugs in the target component. By default, \tool treats failures as test issues (e.g., a component not properly set up) and attempts to fix them. Users can also manually inspect failing tests if they suspect the failure originates from bugs in the target component. The final test suite is executed for coverage and mutation analysis to evaluate adequacy and fault-detection capability.

\noindent
\textbf{Prompt design:}
The repair prompt includes the error message along with the source code of the target component and the generated tests.
It explicitly forbids explanations or comments and requires the LLM to use a significant change in approach on repeated attempts. These constraints reduce verbosity and focus the LLM on producing concise and executable tests.

\begin{table*}[t]
	\centering

	\caption{Evaluation dataset of our experiments. The table reports the details on the target components (CUTs) and their projects.}
    	\begin{scriptsize}
	\label{tab:cut-metrics}

	\rowcolors{2}{gray!10}{white}
 
	\setlength{\tabcolsep}{2pt}
	\renewcommand{\arraystretch}{0.90}
	\resizebox{\linewidth}{!}{%
	\begin{tabular}{lllrrrrrrrrrc}
       \hiderowcolors
		\toprule
		\multirow{2}{*}{\textbf{\makecell{CUT \\ ID}}} 
		    & \multirow{2}{*}{\textbf{Name of CUT}} 
		    & \multirow{2}{*}{\textbf{Project Name}} 
            & \multirow{2}{*}{\textbf{\makecell{Project \\ LOC}}} 
		    & \multirow{2}{*}{\textbf{\makecell{CUT \\ LOC}}} 
		    & \multirow{2}{*}{\textbf{Methods}} 
		    & \multirow{2}{*}{\textbf{Max CC}} 
		    & \multicolumn{2}{c}{\textbf{Tests w/ TDs}} 
            & \multirow{2}{*}{\textbf{Ss}} 
		    & \multirow{2}{*}{\textbf{VOs}} 
		    & \multirow{2}{*}{\textbf{\makecell{Dev \\ Tests?}}} \\
		\cmidrule(l{1pt}r{1pt}){8-9}
		    & & & & & & & \textbf{w/ Ss} & \textbf{w/ VOs} & & & \\
		\midrule
        \showrowcolors
		CAS & \texttt{CustomerApplicationService}  & \href{https://github.com/humank/genai-demo/tree/ecbd93e73cff760bd987e3811028d65b50f82f06}{\underline{E-Commerce Platform}} & 45.9K & 209 & 13 & 9  & 5  & 5  & 8 & 9 & \xmark \\
RRA & \texttt{ResourceRightSizingAnalyzer} & \href{https://github.com/humank/genai-demo}{\underline{E-Commerce Platform}} & 45.9K & 336 & 15 & 13 & 4  & 0  & 6 & 0 & \cmark \\
MAS & \texttt{McpAsyncServer}              & \href{https://github.com/modelcontextprotocol/java-sdk}{\underline{MCP Java SDK}} & 27.4K & 100 & 8  & 13 & 20 & 11 & 23 & 12 & \xmark \\
MUT & \texttt{McpUriTemplateValidator}     & \href{https://github.com/modelcontextprotocol/java-sdk}{\underline{MCP Java SDK}} & 27.4K & 271 & 16 & 10 & 8  & 5 & 16 & 6 & \xmark \\
MQS & \texttt{MaestroQueueSystem}          & \href{https://github.com/Netflix/maestro}{\underline{Maestro}} & 85.7K & 125 & 6  & 8  & 1 & 15  & 1 & 16 & \cmark \\
MPC & \texttt{ModulePermissionConverter}   & \href{https://github.com/Hinadt-Inc/miaocha}{\underline{Miaocha}} & 37.0K & 205 & 8  & 12 & 2  & 2  & 4 & 4 & \xmark \\
QPC & \texttt{QueryPermissionChecker}      & \href{https://github.com/Hinadt-Inc/miaocha}{\underline{Miaocha}} & 37.0K & 150 & 6  & 6  & 5  & 5  & 5 & 5 & \cmark \\
SMS & \texttt{SemanticSchema}              & \href{https://github.com/tencentmusic/supersonic}{\underline{SuperSonic}} & 61.5K & 130 & 19 & 7  & 1  & 0  & 2 & 0 & \cmark \\
ALR & \texttt{AopLogRepository}            & \href{https://github.com/ccmjga/zhilu-admin}{\underline{ZhiLu AI Management}} & 7.5K & 92  & 7  & 22 & 5  & 5  & 5 & 5 & \cmark \\
URE & \texttt{UserRepository}              & \href{https://github.com/ccmjga/zhilu-admin}{\underline{ZhiLu AI Management}} & 7.5K & 97  & 7  & 10 & 3  & 3  & 3 & 3 & \xmark \\
		\bottomrule
	\end{tabular}
	}
    \end{scriptsize}

\end{table*}

\section{Evaluation}
\label{sec:eval}

This section presents the evaluation of \tool{}, which is based on the following research questions (RQs):
\begin{compactitem}
\item[\textbf{RQ1:}] \textbf{Effectiveness} -- \textit{To what extent can \tool generate effective tests?}
\item[\textbf{RQ2:}] \textbf{Complementarity} -- \textit{To what extent do tests generated by \tool complement those generated by baseline approaches and existing tests?}
\item[\textbf{RQ3:}] \textbf{Cost} -- \textit{What is the cost of generating tests using \tool?}
\end{compactitem}

To evaluate \tool{}, we implemented it in a prototype tool for \java projects that use \junit~\cite{junit5} and \mockito~\cite{mockito}. It supports both \maven\ and \gradle-based projects for automated compilation and test execution. We implemented the static analysis and mock extraction of \tool using the \textsc{JavaParser} (\url{https://javaparser.org/}) library, which enables AST-based analyses of source and test files.
All model calls and repair iterations are logged to facilitate reproducibility and further analysis.

\subsection{Dataset}

We evaluated \tool{} on open-source repositories selected from GitHub based on the following criteria. The repository uses at least \java~8, \junit~5, and \mockito~4 to ensure compatibility with modern testing practices. To mitigate data leakage, we included repositories updated after the latest knowledge cutoff date of the model considered (January 2025). The repository has at least 20 GitHub stars, indicating community interest and maintenance. The repository uses \maven or \gradle to facilitate automated building, dependency resolution, and test execution. The repository contains at least one mocked class.

Mining GitHub using the above criteria yielded 24 candidate repositories. We manually examined each repository to identify target components (which we also refer to as components under test or CUTs) based on the following criteria. The CUT has at least 50 lines of code, ensuring sufficient implementation complexity. The CUT implements at least five methods, with at least one having cyclomatic complexity $\ge$ 5, as done in related work~\cite{jahangirova2023sbfttrack}. The CUT is replaced by a test double with stubbings or verify operations in other tests, allowing \tool{} to extract and reuse its defined mocking behavior.

This selection methodology yielded ten CUTs from six unique repositories. Table~\ref{tab:cut-metrics} provides details on the projects and the CUTs. Column ``Max CC'' denotes the highest cyclomatic complexity of the methods in the CUT. Columns under ``Tests w/ TDs'' report the number of tests that define a stubbing (``w/ S'') or a verify operation (``w/ VO) on the test doubles replacing the CUTs. Column ``Dev Tests'' indicates whether the repository contains developer-written tests that directly test (i.e., exercise) the CUTs.

\subsection{Methodology}

\newcommand{\fsmock}[0]{\tool{}\xspace}
\newcommand{\fsnomock}[0]{\textsc{LLM}\xspace}
\newcommand{\zsmock}[0]{\tool{}\(_{ZS}\)\@\xspace}
\newcommand{\zsnomock}[0]{\(\textsc{Baseline}_{ZS}\)\@\xspace}

\noindent\textbf{Baselines}. To evaluate the complementarity of \tool, we considered three baselines: an LLM-based test generation approach (which we refer to as \fsnomock), \randoop~\cite{pacheco2007randoop} (i.e., an approach based on random testing), and the developer-written tests. \fsnomock uses the same prompt and few-shot examples as \tool but omits mock and stub information. This configuration allows for a controlled comparison that isolates the relative contributions of mock information in the LLM-driven test generation process. Although \evosuite{} is generally more effective than \randoop{}~\cite{jahangirova2023sbfttrack}, we used \randoop{} because it provided better support for the recent \java{} versions associated with the projects considered. Across projects, developer-written tests were found for only five CUTs.

\begin{table}[t]
	\centering
	\small
    \vspace{0pt}
	\caption{LLMs used in the evaluation and token costs.}
	\label{tab:llms}
    \vspace{0pt}
	\rowcolors{2}{gray!10}{white}
	\setlength{\tabcolsep}{5pt}
	\renewcommand{\arraystretch}{0.9}
	\resizebox{\linewidth}{!}{%
	\begin{tabular}{llrr}
		\toprule
		\textbf{Provider} & \textbf{Model} & \textbf{Input Cost} & \textbf{Output Cost} \\
		\midrule
		\textsc{OpenAI}    & \gptfouromini & \$0.15 / 1M tok & \$0.60 / 1M tok \\
		\textsc{OpenAI}    & \gptfivemini & \$0.25 / 1M tok & \$2.00 / 1M tok \\
		\textsc{OpenAI}    & \gptfive             & \$1.25 / 1M tok & \$10.00 / 1M tok \\
		\textsc{Anthropic} & \claude              & \$3.00 / 1M tok & \$15.00 / 1M tok \\
		\bottomrule
	\end{tabular}
	}
    \vspace{0pt}
\end{table}

\noindent\textbf{Models}. In RQ1, we evaluated \tool in combination with four LLMs. Table~\ref{tab:llms} lists the models we used and the token costs used at the time we ran the experiments (September 2025). At that time, \gptfouromini{} was recognized as a lightweight, cost-efficient model, \gptfivemini{} and \gptfive{} represented newer reasoning-capable variants, and \claude{} provided a cross-provider comparison. All models were used with default parameters.
In RQ2, we used \gptfivemini{}, as it was a top-performing model in RQ1 and it was cheaper than the other top-performing model \claude{}. Selecting this model allowed us to focus the comparison on the relative benefits of \tool{} while controlling experimental costs and maintaining consistency across runs.

\noindent\textbf{Metrics}. We considered four classes of metrics: \emph{generation}, \emph{compilation}, \emph{execution}, and \emph{quality}. In terms of \emph{generation}, we report the average number of \emph{TeSt generaTed per CUTs} (TST). \emph{Compilation} metrics assess whether \tool produces tests that can be built automatically. These metrics capture feasibility boundaries: percentage of tests that \emph{Compiles on the First Try} (CFT) demonstrates immediate usability, while percentage of tests that \emph{Compiles EVentually} (CEV) reflects the effectiveness of \tool{}'s repair loop. \emph{Execution} metrics measure how many executable tests \tool{} generates. \emph{TestS Passed} (TSP) reports the percentage of tests that pass. \emph{Quality} metrics assess how effectively the generated tests detect faults and exercise code, since quantity and executability alone do not ensure usefulness. We report the minimum (MIN), median (MED), maximum (MAX), and standard deviation (STDEV) of mutation score (fault detection) and line coverage (structural coverage)~\cite{terragni-icpc-2020}. We use PIT~\cite{pitest} (with all mutations enabled) and \textsc{JaCoCo}~\cite{jacoco} to measure mutation score and line coverage, respectively. For fault detection,
we also report \emph{Unique Mutations Killed}, which isolates the exclusive contributions of a given technique, revealing complementarity that aggregate percentages can hide. For structural coverage, \emph{Unique Lines Covered} highlights lines reached only by a technique. Together, these metrics provide a nuanced picture, as it is possible to identify when approaches add non-overlapping value relative to others. 

\noindent\textbf{Setup}. We executed \tool{} and \fsnomock{} ten times to account for randomness in LLM outputs. We repeated the same procedure for \randoop.


\subsection{Results}

\begin{table*}[t]
	\centering
	\begin{scriptsize}
	\caption{Evaluation results obtained by running \tool on the CUTs of the dataset (RQ1).}
	\label{tab:rq1-metrics}
    \vspace{-2mm}
	\rowcolors{2}{gray!10}{white}
	\setlength{\tabcolsep}{4pt}
	\renewcommand{\arraystretch}{0.85}
	\resizebox{\linewidth}{!}{%
	\begin{tabular}{lrrrrrrrrrrrr}
		\toprule
		\multirow{2}{*}{\textbf{Model}} 
            & \multicolumn{1}{c}{\textbf{Generation}} 
			& \multicolumn{2}{c}{\textbf{Compilation}} 
			& \multicolumn{1}{c}{\textbf{Execution}} 
			& \multicolumn{4}{c}{\textbf{Mutation Score (\%)}} 
			& \multicolumn{4}{c}{\textbf{Line Coverage (\%)}} \\
		\cmidrule(l{1pt}r{1pt}){2-2}
        \cmidrule(l{1pt}r{1pt}){3-4}
		\cmidrule(l{1pt}r{1pt}){5-5}
		\cmidrule(l{1pt}r{1pt}){6-9}
		\cmidrule(l{1pt}r{1pt}){10-13}
        & \textbf{TST}
		& \textbf{CFT (\%)} & \textbf{CEV (\%)}
		& \textbf{TSP (\%)}
		& \textbf{MIN} & \textbf{MED} & \textbf{MAX} & \textbf{STDEV}
		& \textbf{MIN} & \textbf{MED} & \textbf{MAX} & \textbf{STDEV} \\
		\midrule
        \gptfouromini{} & 8.5 & 55 & 92 & 81.6 & 0 & 43 & 85 & 27.0 & 0 & 58 & 93 & 33.4\\
        \gptfive{}      & 13.2 & 43 & 100 & 98.6 & 16 & 62 & 100 & 26.7 & 79 & 91 & 100 & 6.4\\
		\gptfivemini{}  & 11.4 & 88 & 100 & 99.7 & 3 & 84 & 100 & 25.6 & 27 & 93 & 100 & 11.4\\
        \claude{} & 45.7 & 44 & 100 & 99.7 & 5 & 89 & 100 & 33.1 & 39 & 94 & 100 & 12.7\\
		\bottomrule
	\end{tabular}
	}
	\end{scriptsize}
    \vspace{-5pt}
\end{table*}

\subsubsection{\textbf{RQ1: Effectiveness}} -- \textit{To what extent can \tool generate effective tests?} Table~\ref{tab:rq1-metrics} reports the results associated with RQ1. The average number of generated tests (TST) varies markedly across models. While \gptfouromini{}, \gptfivemini{}, and \gptfive{} generate between 8.5 and 13.2 tests on average, \claude{} produces a much larger number of tests (45.7 on average). However, this larger test volume does not translate into clearly better quality than other models in terms of median mutation score and line coverage.

\tool is generally successful at producing compilable tests across all considered models.
Although first-try compilation (CFT) varies across models, eventual compilation (CEV) is consistently high, ranging from 92\% for \gptfouromini{} to 100\% for the remainder.
This result indicates that the generation-and-repair loop of \tool is effective at overcoming a substantial portion of the issues that arise during initial test generation.

The execution results further support the effectiveness of \tool. In terms of pass rate, all models except \gptfouromini{} achieve near-perfect results, with TSP values between 98.6\% and 99.7\%.
Taken together, these results indicate that \tool is able to generate not only compilable but also executable tests with high reliability, especially when paired with \gptfivemini{}, \gptfive{}, or \claude{}.

The mutation score results show that \tool can generate tests with strong fault-detection capability.
All models reach high maximum mutation scores, ranging from 85\% for \gptfouromini{} to 100\% for \gptfive{}, \gptfivemini{}, and \claude{}. The median mutation score provides a more robust picture of typical performance across runs. Here, \gptfivemini{} and \claude{} obtain the strongest medians, with 84\% and 89\%, respectively, while \gptfive{} reaches 62\% and \gptfouromini{} 43\%. At the same time, the standard deviation values are relatively large for all models, indicating variability across runs and CUTs. Still, the high medians and maxima for \gptfivemini{}, \gptfive{}, and \claude{} show that \tool can often generate tests that kill a substantial fraction of mutants. Among these models, \gptfivemini{} stands out because it combines a high median mutation score with much stronger compilation and execution behavior than \claude{} (and at much lower cost, see RQ3).

The line coverage results further support the effectiveness of \tool.
Median line coverage is high for the three strongest models: 91\% for \gptfive{}, 93\% for \gptfivemini{}, and 94\% for \claude{}.
Moreover, all three models reach 100\% maximum line coverage, while \gptfouromini{} reaches a maximum of 93\%.
\gptfouromini{} shows substantially weaker typical performance, with a median line coverage of 58\%, whereas the other three models consistently exercise a large portion of the target code.
Although \claude{} attains the highest median line coverage, its advantage over \gptfivemini{} is marginal (94\% vs.\ 93\%).
Thus, considering line coverage together with compilation and execution reliability, \gptfivemini{} emerges as the most practical and effective choice.

Overall, the results provide evidence that \tool is effective at generating useful tests.
Across the stronger models, \tool almost always produces executable tests after repair and often achieves high line coverage and mutation scores, indicating that the generated tests are not merely runnable but also meaningful from a testing perspective.
Among the evaluated models, \gptfivemini{} offers the best overall trade-off.
It achieves the highest first-try compilation rate, perfect eventual compilation, near-perfect pass rate, and strong median mutation score and line coverage.
At the same time, it is considerably smaller and cheaper than \gptfive{} and \claude{}, making it the most effective and practical model for \tool based on our evaluation.

\vspace{5pt}
\noindent\subsubsection{\textbf{RQ2: Complementarity} -- \textit{To what extent do tests generated by \tool complement those generated by baseline approaches and existing tests?}}
Across CUTs, \tool achieves higher unique mutation kill rates than both \fsnomock and \randoop only (shown in Table~\ref{tab:baseline-comparison}). CAS and RRA are prime examples of this, showing a \tool detection rate of 24.56\% and 25.64\%, compared to \fsnomock (~0-12.82\%). \tool achieves a strictly higher unique mutation killed rate than \fsnomock for 40\% of CUTs and \randoop for 50\% of CUTs. \randoop rarely kills any unique mutations, with killing 2.56\% only on RRA. For MAS, MUT, QPC, and SMS, the table shows 0\% across all approaches, indicating that \tool{} performs similarly across all cases. This suggests that mock information generally helps identify unique mutation cases missed by other tools. The low \randoop performance indicates that a simple traditional test generation approach might struggle to detect nuanced faults without the contextual and semantic reasoning available to LLMs.
\begin{table*}[t]
	\centering
	\small
	\caption{Baseline Comparison with \fsnomock, \randoop{} and developer-written tests by CUTs (RQ2)}
	\label{tab:baseline-comparison}
    \vspace{0pt}
	\rowcolors{2}{gray!10}{white}
	\setlength{\tabcolsep}{11pt}
	\renewcommand{\arraystretch}{0.80}
	\resizebox{\linewidth}{!}{%
	\begin{tabular}{lrrrrrrrr}
		\toprule
		\textbf{CUT ID} 
		    & \multicolumn{4}{c}{\textbf{Unique Mutations Killed (\%)}} 
		    & \multicolumn{4}{c}{\textbf{Unique Lines Covered (\%)}} \\
		\cmidrule(l{1pt}r{1pt}){2-5}
		\cmidrule(l{1pt}r{1pt}){6-9}
		    & \textbf{\fsmock} 
		    & \textbf{\fsnomock} 
		    & \textbf{\randoop{}} 
		    & \textbf{Dev Tests} 
		    & \textbf{\fsmock} 
		    & \textbf{\fsnomock} 
		    & \textbf{\randoop{}} 
		    & \textbf{Dev Tests} \\
		\midrule
		CAS & 24.56 & 1.75 & 0.00 & -- & 0.00 & 1.18 & 0.00 & -- \\
		RRA & 25.64 & 12.82 & 2.56 & 2.56 & 2.61 & 0.65 & 2.61 & 2.61 \\
		MAS & 0.00 & 0.00 & 0.00 & -- & 0.00 & 0.00 & 0.00 & -- \\
		MUT & 0.00 & 0.00 & 0.00 & -- & 0.00 & 0.00 & 0.00 & -- \\
		MQS & 10.53 & 5.26 & 0.00 & 0.00 & 1.56 & 0.00 & 0.00 & 0.00 \\
		MPC & 1.23 & 0.00 & 0.00 & -- & 0.83 & 0.00 & 0.00 & -- \\
		QPC & 0.00 & 0.00 & 0.00 & 0.00 & 0.00 & 0.00 & 0.00 & 0.00 \\
		SMS & 0.00 & 0.00 & 0.00 & 0.00 & 0.00 & 0.00 & 0.00 & 0.00 \\
		ALR & 4.26 & 4.26 & 0.00 & 0.00 & 0.00 & 0.00 & 0.00 & 0.00 \\
		URE & 0.00 & 0.00 & 0.00 & -- & 1.79 & 0.00 & 0.00 & -- \\
		\bottomrule
	\end{tabular}
	}
    \vspace{-10pt}
\end{table*}
Unique line coverage is generally low across all CUTs, with only minor contributions from \tool (e.g., 2.61\% for \textsc{RRA}, 1.56\% for \textsc{MQS}, and 1.79\% for \textsc{URE}) and one case for \randoop (2.61\% for \textsc{RRA}).
Mock information yields a small but measurable advantage, exposing a few additional lines not reached by other techniques. However, its overall impact on coverage is less pronounced than on mutant detection.

Developer tests contribute the fewest unique mutations (2.56\% for \textsc{RRA}, none elsewhere), indicating that LLM-generated tests can reveal faults missed by human developers.
Mock information enhances this advantage by improving fault detection diversity. For line coverage, differences are minor.
\tool-generated tests again show the highest exclusive coverage (e.g., 2.61\% for \textsc{RRA}, 1.56\% for \textsc{MQS}), while no-mock and developer tests contribute little additional coverage. Although the differences are minor, there is evidence that mock information can help LLMs reach certain lines missed by \fsnomock, \randoop, and developer tests.

\begin{table}[t]
    \centering
    \vspace{0pt}
    \caption{Average test generation cost and token usage (RQ3)}
    \vspace{0pt}
    \label{tab:avg-test-cost}
    \rowcolors{3}{gray!10}{white}

    \setlength{\tabcolsep}{3pt}
    \resizebox{\linewidth}{!}{%
    \begin{tabular}{lrrr|rrr}
        \hiderowcolors
        \toprule
         & \multicolumn{3}{c|}{\textbf{\tool}} & \multicolumn{3}{c}{\textbf{\fsnomock}} \\
        \cmidrule(r){2-4} \cmidrule(l){5-7}
       \textbf{Model} & \textbf{Cost} & \textbf{Input } & \textbf{Output} & \textbf{Cost} & \textbf{Input} & \textbf{Output} \\
& \textbf{USD} & \textbf{Tokens} & \textbf{Tokens} & \textbf{USD} & \textbf{Tokens} & \textbf{Tokens} \\
       \showrowcolors
        \midrule
        \gptfouromini   & \$0.0076 & 36,761 & 3,506  & \$0.0062 & 26,805 & 3,632  \\
        \gptfivemini    & \$0.0176 & 16,477 & 6,723  & \$0.0171 & 13,352 & 6,852  \\
        \gptfive        & \$0.1028 & 15,228 & 8,377  & \$0.0786 & 9,043  & 6,733  \\
        \claude         & \$0.5132 & 78,713 & 18,471 & \$0.4358 & 58,918 & 17,272 \\
        \bottomrule
    \end{tabular}
    }
\end{table}
\vspace{5pt}
\subsubsection{\textbf{RQ3: Cost} -- \textit{What is the cost of generating tests using \tool?}}

Table~\ref{tab:avg-test-cost} presents the average test generation cost (USD) and input/output token usage for \tool and \fsnomock across LLMs. Costs were derived from the input and output tokens consumed in each run and the corresponding model-specific API pricing.

At the prompting level, providing mock information (\tool vs. \fsnomock) slightly increases input tokens and thus cost by increasing the prompt length. These effects typically add around 5-15\% of cost within the same model, and start to be noticeable with models where token rates are higher.

At the model level, cost increases predictability with model size and capabilities. \claude has the highest cost, followed by \gptfive (\verb|~|\$0.08--\$0.11), whereas the mini variants are quite efficient (typically under \$0.02), with \gptfivemini{} being the most cost-effective model in terms of cost and quality of generated tests.



\section{Discussion}

\noindent
\textbf{Interpretation of Results.}
The results indicate that \tool{} can guide LLM models toward useful tests.
By leveraging mock information, \tool{} provides contextual cues that help uncover behaviors and faults missed by a vanilla LLM, \randoop{}, or developer-written tests.
This is supported by the moderate number of unique mutations killed and unique lines of code covered by \tool{}.
Because each experiment was repeated ten times, we are confident that the observed benefits are consistent and not due to random variation in LLM outputs.
Mock-informed test generation consistently killed mutations that were otherwise undetected. This suggests that stubbings and verify operations activate distinct reasoning pathways within models and help them focus on different faults.

\noindent
\textbf{Complementary.} We applied a \emph{Kruskal--Wallis test} to assess whether mutation scores and line coverage differ significantly across techniques.  All \(H\) values (0.19--1.94) and \(p\)-values (\(\geq 0.584\)) indicate no statistically significant differences between \tool{} and the \textsc{Baseline}. As such, \tool{} complements existing approaches. It can enhance existing test suites by generating additional tests that detect more faults and expand behavioral coverage.
Indeed, the contextual cues derived from stubbing and verify operations that guide test generation may reduce test diversity, since the model tends to follow the provided interaction patterns. Future LLM-driven test generation could consider combining prompts both with and without mock information when available.

\noindent
\textbf{Cost and Practical Implications.}
The cost analysis shows that \tool{} incurs only a modest cost overhead (about 5--15\% higher than LLM generation without mock information) while killing unique bugs and covering additional code. 
This trade-off is favorable for practical adoption, as \tool{} can deliver measurable testing benefits at an affordable additional computational cost.

\subsection{Threats to Validity}

\noindent
\textbf{Generalizability.}
We evaluated our approach on ten classes from six projects. Future work is needed to assess whether the approach generalizes across other projects, languages, and mocking frameworks. Although the current prototype and experiments focus on \mockito\ and \java-based projects, the approach can be extended to other languages and frameworks. For example, to \textsc{Python} projects using \textsc{unittest.mock} or \textsc{JavaScript} projects using \textsc{Jest}. 

\noindent
\textbf{Metric limitations.}
The metrics we used (generation, compilation, execution, and coverage) capture only part of test quality. Generated tests that compile and pass may still need improvement. Future work should include a manual assessment of the quality of generated tests.

\noindent
\noindent
\textbf{Model bias and training data leakage.}
To mitigate data leakage, we selected projects updated after the training cut-off date of the selected models. Despite this precaution, it remains difficult to guarantee that the results stem purely from reasoning rather than memorization of data (possibly from similar projects) seen during training.

\section{Related Work}

\noindent\textbf{Traditional Test Generation Techniques.}
Random testing~\cite{pacheco2007randoop} and search-based software testing~\cite{fraser2011evosuite} reduce manual effort by automatically producing tests to maximize structural coverage.
Similar to \tool{}, these techniques leverage information in source code to enhance test generation.
\textsc{MSeqGen}~\cite{DBLP:conf/sigsoft/ThummalapentaXTHS09} mines method call sequences from code repository to achieve higher coverage.
Fraser and Zeller~\cite{DBLP:conf/icst/FraserZ11} mine object usage patterns from existing code to generate meaningful tests.
Our work complements them by utilizing mocking information to guide test generation toward better tests.


\noindent\textbf{Automated Test Generation with LLMs.}
LLM-based approaches generate unit tests via prompt engineering and structured context. Studies on zero-shot, few-shot, and Chain-of-Thought prompting show that richer, code-aware inputs (e.g., signatures, existing tests, stack traces) outperform plain natural language prompts~\cite{ouedraogo2024large}. Accordingly, pipelines typically collect project context, query an LLM, and iteratively compile/execute and repair failing outputs~\cite{schafer2023empirical,wang2024software,ravi2025llmloop,terragni2025future}. Similar to prior work, \tool relies on prompt design and post-processing. However, unlike approaches that build context solely from source code and project metadata, \tool also leverages developer-written mocks and stubs as structured input to guide the LLM in generating realistic tests for mocked and stubbed classes. Although prior tools may generate tests with test doubles, none explicitly exploit existing mocking information to guide test generation. To the best of our knowledge, \tool is the first to do so.

\noindent\textbf{Test Doubles in Software Testing.}
Recent work has introduced tools that analyze, generate, or refactor mocks and stubs to improve testing quality and maintainability. 
\textsc{MockSniffer}~\cite{zhu2020mocksniffer} characterizes recommends mocking decisions.
Empirical studies on mock assertions~\cite{zhu2025understanding} suggest that mocks capture developers' behavioral intent and domain knowledge.
\tool directly utilize this encoded intent to inform LLM-driven test generation.
\textsc{StubCoder}~\cite{zhu2023stubcoder} generates and repairs stub code via evolutionary search to keep tests passing as production code evolves.
RICK~\cite{tiwari2024mimicking} records production executions and generates tests that mimic observed behavior using test doubles. ARUS~\cite{li2024automatically} improves maintainability by detecting and removing unnecessary stubbings.
Compared with these approaches, \tool does not aim to generate or improve test doubles but instead leverages the behavioral information already encoded in them to guide LLMs in generating new tests.

\section{Conclusions and Future Work}

This paper introduced \tool, the first LLM-based test generation approach that leverages developer-written mocking information to guide test creation. \tool generates tests that uncover lines of code and mutants missed by baseline approaches, showing that mock information provides valuable contextual cues for producing realistic and useful tests. The approach complements developer-written suites, traditional tools such as \randoop, and an vanilla LLM-based approach. The cost overhead of 5–15\% with respect to the LLM baseline considered is acceptable given the corresponding improvements in test quality and fault detection capability.

Our findings suggest that future LLM-driven test generation should incorporate available mocking information as contextual guidance to produce more comprehensive and realistic tests. Rather than replacing existing methods, mock-informed generation should complement them within a holistic LLM-based testing framework that integrates both mock-informed and non–mock-informed prompts.

As the first work of its kind, \tool{} opens several promising research directions. Future work includes ablation studies to assess the relative impact of mocking information, automatic extraction of project-specific few-shot examples to improve performance, dynamic collection of mocking data to capture richer behaviors, and establishing traceability links between test doubles and generated tests to better understand their influence on test creation.

\balance
\bibliographystyle{IEEEtran}
\bibliography{IEEEabrv,bib}

\end{document}